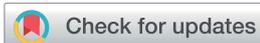

Check for updates

# Bridgman method grown Cs₂Li₃I₅: an inter-alkali metal scintillator with high lithium content

Kateřina Křehlíková, [ID] *[ab] Vojtěch Vaněček, [ID] [a] Robert Král,[a] Romana Kučerková,[a] Petra Zemenová,[a] Jan Rohlíček,[a] Petr Průša [ID] [ac] and Kateřina Rubešová[b]

In this study, we report on the growth of ternary caesium lithium iodide ($Cs_2Li_3I_5$, CLI) bulk crystals, both undoped and doped with thallium (Tl) and indium (In), using the miniaturised vertical Bridgman method (mVB). X-ray Powder Diffraction (XRPD) confirmed the presence of the ternary CLI phase in all three crystals, with CLI : In appearing homogeneous structure-wise throughout the entire ingot. Measurements of radioluminescence (RL), photoluminescence emission (PL), photoluminescence excitation (PLE) spectra, and photoluminescence decay kinetics (PL decay) demonstrated that the primary luminescence centers originate from the matrix itself. When doped with thallium, the efficiency of the luminescence was significantly increased. Furthermore, CLI : Tl and CLI : In crystals exhibited emission spectra similar to those of their doped caesium iodide counterparts, CsI : Tl and CsI : In, respectively. The main component of the PL decay was 523 ns, 557 ns, and 554 ns for the undoped, Tl⁺-doped, and In⁺-doped crystals, respectively. It is worthy of note that only CLI : Tl exhibited a single exponential decay. Differential Scanning Calorimetry (DSC) measurements revealed two endothermic peaks corresponding to the eutectic and liquidus temperature for CLI and CLI : In, indicating that this ternary compound has a congruent melting behaviour. Finally, the melting point of CLI was estimated to be approximately 220 °C.

## Introduction

Neutrons and their detectors play a crucial role in a variety of applications, including nuclear power generation, nuclear decommissioning and decontamination, border and homeland security, nuclear non-proliferation, neutronography, and medicine.[1–5] Several nuclear reactions and interactions are employed for neutron detection, including ³He(n,p)³H, ¹⁰B(n,α)⁷Li, ⁶Li(n,α)³H, elastic scattering, and fission reactions. Each method has its own advantages and disadvantages. For instance, ³He has been costly due to a global shortage since 2008,[3,4,6,7] and elastic scattering is only suitable for fast neutrons.

The ⁶Li(n,α)³H reaction has a slightly lower cross-section of 941 barns at an energy of 25 meV.[8] However, its relatively high energy reaction $Q$ value of 4.78 MeV makes it suitable for amplitude n/γ (neutron/gamma) discrimination. Furthermore, the reaction produces the heavy charged particles required for pulse shape n/γ discrimination. Finally, the existence of a neutron-insensitive stable ⁷Li isotope enables the construction of a complementary detector that can correct and subtract responses induced by gamma radiation (γ-ray).

[a]*Institute of Physics of the Czech Academy of Sciences, Cukrovarnicka 10/112, 162 00 Prague, Czech Republic*

[b]*Faculty of Chemical Technology, University of Chemistry and Technology, Technicka 5, 166 28 Prague, Czech Republic*

[c]*Faculty of Nuclear Sciences and Physical Engineering, Czech Technical University in Prague, Brehova 78/7, 115 19 Prague, Czech Republic*

Ternary alkali halides, such as caesium lithium iodide ($Cs_2Li_3I_5$, CLI), could represent an alternative material for the detection of neutrons, X-, and γ-rays by a single detector, *e.g.*, for mixed radiation field monitoring or a combined X-ray radiography and neutronography imaging device. According to the CsI–LiI phase diagram reported by Sangster and Pelton,[9] no ternary phase is present; however, the formation of a ternary CLI phase was reported by Meyer and Gaebell already in 1983,[10] and basic crystallographic properties were measured (see Table 1). Later, CLI thin films were studied by Nagarkar *et al.*[6] According to this study, CLI doped with Eu²⁺ exhibited a light yield of *ca.* 40 000 to 55 000 photons/neutron with emission peaking at *ca.* 450 nm. The significant difference in the primary decay time for γ (500 ns) and neutron (250 ns) excitation enabled a facile pulse shape discrimination (PSD).[6] PSD capability with sufficient figure of merit (FOM) is crucial for applying a scintillator in mixed n/γ radiation fields to achieve sufficient γ-ray rejection,[6] which is always desired.[11] Nevertheless, the preparation of large CLI single crystals and the study of their thermal properties have not yet been reported.

**Table 1** The summary of crystallographic properties of $Cs_2Li_3I_5$ reported by Meyer and Gaebell.[10]

| Crystal structure | Monoclinic |
| --- | --- |
| Lattice parameters | $a = 16.668(6)$ Å |
|  | $b = 4.721(1)$ Å |
|  | $c = 10.987(4)$ Å |
|  | $\beta = 115.73(3)°$ |
| Space group | $C2/m$ (no. 12) |
| Cell volume | 778.84 Å³ |
| Coordinations of cations | 8(Cs) |
|  | 6(Li1) |
|  | 4(Li2) |

The goal of this work is to prepare single crystals of ternary CLI, confirm their phase purity, and perform basic physical, structural, optical, and thermal characterisations. The aim is also to investigate the influence of doping with monovalent ns$^2$ ions (Tl$^+$, In$^+$) on the luminescence properties of the CLI matrix.

## Experimental

Ternary Cs$_2$Li$_3$I$_5$ (CLI) crystals were prepared from raw materials of caesium iodide (CsI, Monokrystaly Turnov) and commercially available lithium iodide (LiI, 99.95%, Alfa Aesar). The CsI was purified of oxy- and hydroxy-impurities by a combination of vacuum drying, introduction of a gaseous mixture of halogenation agents into the melt, and subsequent zone refining (>20 passes of the melted zone). A detailed description of this purification technique was reported by Nitsch *et al.*[12,13] LiI was not further purified and was used as purchased.

At first, the undoped CLI was prepared by direct synthesis of purified CsI and raw LiI mixed in a stoichiometric ratio of 2 : 3, which was then placed inside a quartz ampoule. Afterward, the materials were purified with the gaseous mixture of halogenation agents introduced into their melt and sealed in the ampoule with a hydrogen–oxygen torch. The CLI synthesised in this way (further denoted as 'Synthesis I') was transported into a glovebox unit (with a controlled nitrogen atmosphere) and ground into a powder in a corundum mortar. The synthesised CLI powder was then sealed in a quartz ampoule under vacuum for subsequent vertical Bridgman growth.

Secondly, the CLI crystals doped with Tl$^+$, and In$^+$ were synthesised using a different approach (further denoted as 'Synthesis II'). In this process, CsI was placed inside a quartz ampoule, purified by a mixture of halogenating agents, and transferred to a glovebox unit, where the LiI powder (in a stoichiometric ratio of 2 : 3 to CsI) and the required amount of dopant were added. As dopants, commercially available TlI (purity 99.95%, Nuvia) and InI (purity 99.95%, Sigma Aldrich), both purified by the mixture of halogenating agents, were used for CLI : Tl and CLI : In, respectively, at a concentration of 0.1 at%. The ampoules containing both compositions were then sealed under vacuum and melted to ensure the proper homogenisation and formation of the ternary CLI phase.

The CLI, CLI : Tl, and CLI : In crystals were grown by the miniaturised vertical Bridgman method (mVB) using a micro-pulling-down apparatus, as described by Vaněček *et al.*[14] The ampoules containing the materials were placed in the hot zone, which consisted of a graphite tube that served as the heating element, forming a single-zone, inductively heated Bridgman furnace. The alumina shielding was not included in the hot zone due to the very low melting point of the CLI compound to achieve a steeper temperature gradient and avoid high undercooling of the melt. Afterward, ampoules were pulled down at a rate of 0.6 mm h$^{-1}$ until complete melt solidification was achieved. The crystals were cooled to room temperature (RT) for 20 hours.

The X-ray powder diffraction (XRPD) data were collected in the angular range (2θ) 4–100° with a 0.013° step using the Debye–Scherrer transmission configuration on the powder diffractometer Empyrean of PANalytical ($\lambda_{Cu,K\alpha}$ = 1.54184 Å) that was equipped with a focusing mirror, a capillary holder, and a PIXcel3D detector. Before the experiments, the diffractometer was aligned using a Si standard. The XRPD patterns were evaluated using the PANalytical X'Pert HighScore Plus 3.0e commercial software. For the XRPD analysis, the grown crystals were ground using a corundum mortar and pestle. The prepared powders were placed in borosilicate-glass capillaries with a diameter of 0.5 mm and sealed with rubber to prevent sample deterioration due to exposure to the air. The diffraction experiment for each sample was made from 4° to 100° 2θ, which was repeated 20 times, with an overall time of approximately 20 h to collect a pattern of acceptable quality. To evaluate phase purity along the growth axis, three samples from each crystal were collected from the front, middle, and tail parts of the grown ingot, further denoted as A, B, and C, respectively.

Photoluminescence emission (PL), photoluminescence excitation (PLE), and radioluminescence (RL) spectra, as well as photoluminescence decay kinetics (PL decay), were measured in the spectral range of 190–800 nm at RT using a custom-made spectrofluorometer 5000M, Horiba Jobin Yvon. The X-ray tube, tungsten (W) anode (40 kV, 15 mA), Seifert GmbH (RL spectra), and steady-state laser-driven xenon lamp, Energetiq EQ-99X LDLS (PL and PLE spectra) were used as the excitation sources, respectively. The nanoLED (295 nm) ns pulse source, IBH Scotland, was used for the excitation of the PL decays. The detection part of the set-up consisted of a single grating monochromator and a photon-counting detector TBX-04, IBH Scotland. Monochromator slit sizes were optimized for each spectroscopic measurement. For RL spectra, a slit size of 8 nm was used. PL and PLE spectra were measured with a slit size of 2/2 nm. PL decay measurements utilized different slit sizes depending on the sample: 32/32 for undoped CLI and CLI : In, and 16/16 nm for CLI : Tl. The measured spectra were corrected for the spectral dependence of the excitation energy (PLE) and the spectral dependence of the detection sensitivity (PL). Due to the high hygroscopicity of the CLI crystals, the samples were covered with UV/VIS-transparent Fluka immersion oil to decelerate the deterioration due to exposure to the air.

Differential scanning calorimetry (DSC) was performed on CLI and CLI : In powder samples using a simultaneous thermal analyser, Themys 2400 (Setaram). The powders (*ca.* 25 mg) were sealed in quartz ampoules under a vacuum and placed in an alumina crucible. The ampoules were measured using heating and cooling rates of 10 K min$^{-1}$ in the temperature range 25–350 °C under an argon flow of 20 mL min$^{-1}$ (purity 5.0 N) with 5- and 4-cycles repetition for CLI and CLI : In, respectively. An alumina crucible containing an empty quartz ampoule sealed under vacuum was used as a reference. The DSC apparatus was calibrated in the 25–1300 °C temperature range using the following standards: In, Sn, Zn, Al, Ag, and Au. The standard deviation of the performed calibrations was ±0.8 K, and the DSC baseline was measured using a second empty ampoule. The measured data were processed using the program Calisto Processing.

Due to the hygroscopic nature of the used materials, great care was taken to prevent degradation through reaction with moisture and oxygen. Therefore, all procedures, including the weighing of starting materials, the feeding of quartz ampoules, the closure of valves, the manufacture of grown crystals (including the cutting and polishing of samples), and the preparation of powder in a corundum mortar, were performed in an MBraun LabStar glovebox unit filled with an inert nitrogen atmosphere and with an oxygen and moisture content below one ppm.

## Results and discussion

The CLI, CLI : Tl, and CLI : In crystals were successfully grown using the mVB method. The undoped crystal is colourless, whereas the Tl[+]- and In[+]-doped crystals are brown (see Fig. 1I–III) due to a different preparation approach (Synthesis II). Samples for optical characterisation were cut and polished to a thickness of 1.5 mm for all three crystals (see Fig. 1IV–VI). Due to multiple cracks in the grown ingots, the samples for optical characterisation are irregular in shape.

The XRPD analysis revealed that all samples collected from parts A–C contained the $Cs_2Li_3I_5$ phase (PDF code: 01-076-1482) with varying purity. CsI, LiI, $Li_2O_2$, and $Li_2O$ were identified as secondary phases. However, the presence of $Li_2O$ is questionable, as its diffraction lines are approximately 0.1° $2\theta$ offset from the measured peaks. In addition to these impurities, A- and B-parts of the CLI and CLI : Tl contained varying amounts of an unidentified phase. The highest concentration of this phase was observed in the B-part of the CLI sample. XRPD results collected from A–C parts of all crystals are summarised in Table 2. However, Fig. 2 only depicts and compares the XRPD patterns of the tail parts (last-to-freeze, *i.e.*, C-part). All patterns were compared with PDF no. 01-076-1482 (shown in dashed lines), and the impurity phases were labelled with an asterisk. Table 2 shows that all CLI, CLI : Tl, and CLI : In crystals contain the $Cs_2Li_3I_5$ ternary phase. CLI : In is the only sample that is a pure $Cs_2Li_3I_5$ monoclinic phase across the A- and B-parts of the ingot, containing only traces of an unknown impurity in the C-part. It is necessary to consider the relatively low sensitivity of XRPD, which only allows impurity phases to be detected down to a concentration of approximately 1 wt%.

**Table 2** The composition of the CLI, CLI : Tl and CLI : In crystals across the ingot (parts A, B, C). PDF used for phase evaluation were: 01-076-1482 for $Cs_2Li_3I_5$, 01-074-6256 for $Li_2O$, 01-071-5197 for CsI, 01-074-0115 for $Li_2O_2$, and 04-016-3486 for LiI. The weight fractions were estimated using reference intensity ratios (RIR) without accounting for the unknown impurity. The amount of the unknown impurity (UI) is qualitatively classified as present (+) or absent (−)

| Crystal | Part | Weight fraction (%) | | | | | UI |
| | | $Cs_2Li_3I_5$ | $Li_2O$ | CsI | $Li_2O_2$ | LiI | |
|---|---|---|---|---|---|---|---|
| CLI | A | 100 | — | — | — | — | + |
| | B | 53 | 34 | 2 | 11 | — | + |
| | C | 84 | 16 | — | — | — | + |
| CLI : Tl | A | 78 | 21 | 2 | — | — | + |
| | B | 100 | — | — | — | — | + |
| | C | 85 | — | — | — | 15 | + |
| CLI : In | A | 100 | — | — | — | — | − |
| | B | 100 | — | — | — | — | − |
| | C | 100 | — | — | — | — | + |

On the contrary, CLI and CLI : Tl crystals were composed primarily of the $Cs_2Li_3I_5$ phase in all parts. Secondary binary phases were identified alongside traces of unknown impurities.

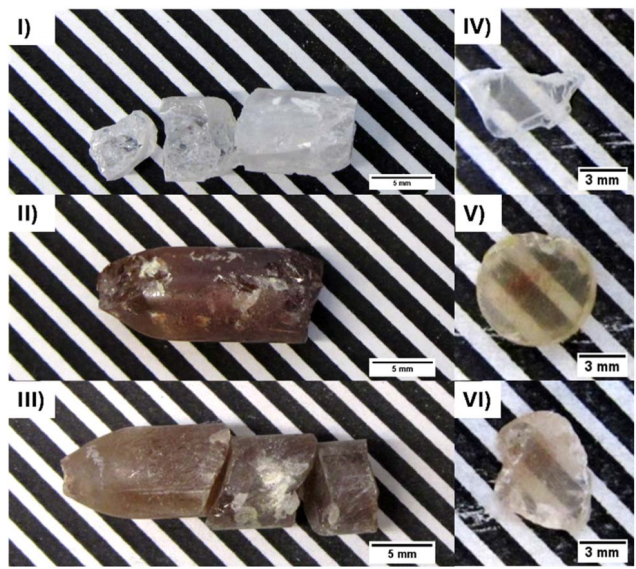

**Fig. 1** Photos of as-grown crystals (I–III) and samples prepared for optical characterization (IV–VI) of undoped CLI (I) and (IV), CLI : Tl (II) and (V), and CLI : In (III) and (VI), respectively.

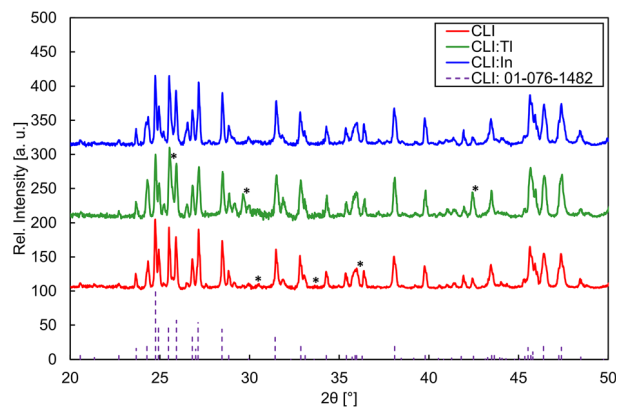

**Fig. 2** XRPD patterns of C-parts of all crystals compared to PDF no. 01-076-1482. The most intense peaks of impurity phases are labelled with an asterisk.

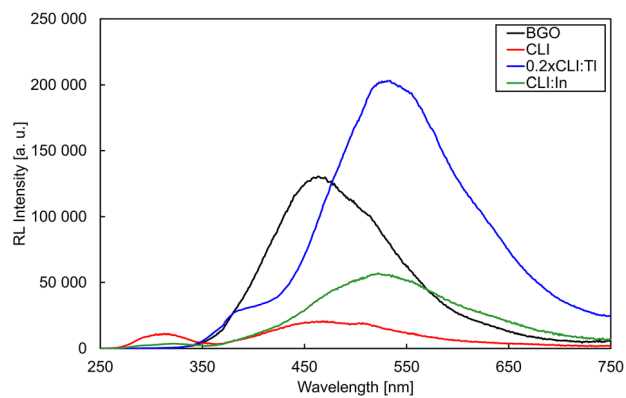

**Fig. 3** The RL spectra of BGO, CLI, CLI : Tl, and CLI : In. The RL spectrum of CLI : Tl was divided by 5 for better clarity.

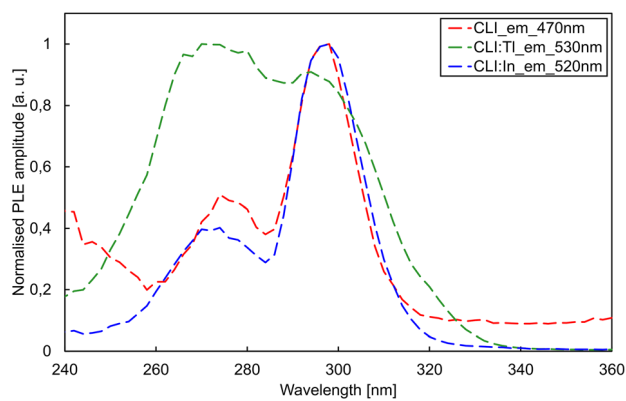

**Fig. 4** PLE spectra of CLI, CLI : Tl, and CLI : In crystals for emissions 470 nm, 530 nm, and 520 nm, respectively.

In the case of undoped CLI, $Li_2O$, CsI, and $Li_2O_2$ were present in the middle and tail parts of the crystal, while CLI : Tl was a pure $Cs_2Li_3I_5$ phase only in the middle part. The front part of CLI : Tl contained $Li_2O$ and CsI phases, whereas the tail part was contaminated by LiI (as the only case of CsI detected). Such a difference in the phase segregation in the prepared crystals is most likely due to the raw LiI not being purified prior to crystal growth. It is important to note that CLI and CLI : Tl (as well as CLI : In) were prepared using different approaches, as described in the 'Experimental' section of this paper. Based on these results, Synthesis I is not suitable, and Synthesis II should be used in the future alongside the purification of the LiI raw material.

The radioluminescence spectra measured for all the CLI, CLI : Tl, and CLI : In crystals, as well as for the bismuth germanate ($Bi_4Ge_3O_{12}$, BGO) crystal used as a reference material, are depicted in Fig. 3. For better clarity, Fig. 3 shows the spectrum of CLI : Tl divided by the factor of five, making it comparable to CLI and CLI : In. The emission maxima and the integral radioluminescence efficiency (integral of spectra in Fig. 3) compared to the BGO are summarized in Table 3. The RL spectra of all the CLI, CLI : Tl, and CLI : In crystals consist of multiple broad emission bands. However, there are significant differences among the samples. The CLI RL spectrum shows two main, distinct, broad peaks with maxima at 313 and 468 nm. The addition of the ns$^2$ dopants resulted in the suppression of the peak at 313 nm, while the second broad peak increased in intensity and red-shifted, resulting in a broad band peaking at 520–535 nm. Such an effect is much more significant for Tl$^+$- than In$^+$-doping, as the RL integral intensity of the CLI : Tl sample reached *ca.* 900% of that of BGO, representing an increase in RL efficiency of over 40 times compared to undoped CLI. Moreover, this is associated with a redshift of the emission band of CLI : Tl at a maximum of 534 nm, which is very similar to the CsI : Tl (530 nm),[15] and of CLI : In with a maximum at 522 nm resembling CsI : In (545 nm).[16] It is important to note that, based on the width and shape of the long-wavelength emission band, it is most probably composed of several overlapping peaks. However, further investigation is necessary to deconvolve the emission band and fully explain its origin.

The photoluminescence excitation (PLE) and emission (PL) spectra are depicted in Fig. 4 and 5, respectively. Based on RL spectra, the monitored emission wavelengths were selected as 470, 530, and 520 nm for CLI, CLI : Tl, and CLI : In, respectively. The PLE spectra (Fig. 4) consist of two overlapping bands, peaking at around 270 and 295 nm for all crystals.

Excitation in the wavelength range of 290–295 nm resulted in a broad band emission with a similar position and shape for all three crystals (see Fig. 5). From these measurements, we can conclude that the luminescence centres of all the crystals have a similar origin and that the main luminescence centre originates from the matrix. Based on the significant Stokes shift of

**Table 3** Summary of CLI, CLI : Tl, and CLI : In RL maxima and RL efficiency related to BGO. Monochromator slit size of 8 nm was used, giving an uncertainty of measurement ± 8 nm

| Crystal | Peak (nm) | RL efficiency (%) |
|---|---|---|
| CLI | 313, 468, 508 | 21 |
| CLI : Tl | 380, 534 | 915 |
| CLI : In | 320, 522 | 53 |

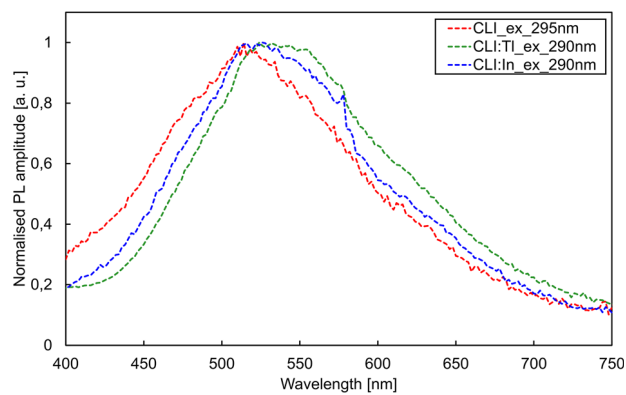

**Fig. 5** PL spectra of CLI, CLI : Tl, and CLI : In crystals for excitations 295 nm, 290 nm, and 290 nm, respectively.

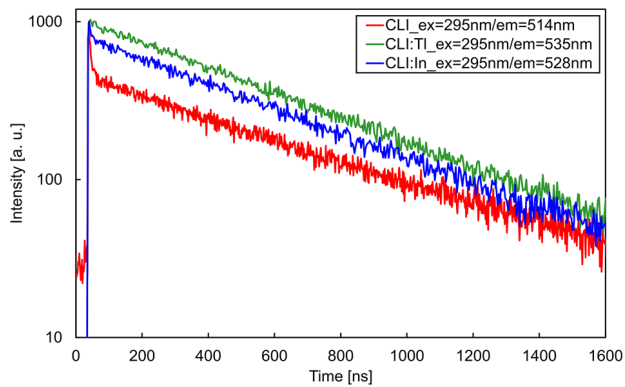

**Fig. 6** PL decays of CLI, CLI : Tl and CLI : In crystals for fixed excitation and emission wavelengths (see legend).

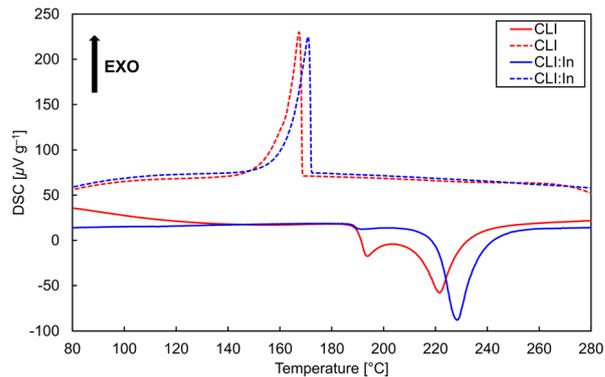

**Fig. 7** DSC heating and cooling (in solid and dashed line, respectively) curves of CLI and CLI : In samples. The standard deviation of performed calibrations was in the range of ±0.8 K.

*ca.* 1.8 eV and the FWHM of the emission band of *ca.* 0.8 eV, we tentatively ascribe the origin of this emission to a trapped exciton (TE). In the undoped CLI, the TE can be formed by either self-trapping or trapping at defects or unintentional impurities in the crystal lattice. In the case of the doped crystals, the origin of the emission is most probably an impurity-(Tl/In)-trapped exciton. The increase in exciton binding energy of the impurity-trapped excitons could explain the redshift in the PL emission band. Adding the $Tl^{+}$ dopant enhances the performance, as can be seen from the RL results in absolute form (Fig. 3).

The PL decay curves of CLI, CLI : Tl, and CLI : In are depicted in Fig. 6, and the fitting parameters are summarised in Table 4. All three crystals exhibit dominant (>98%) components with a decay time of *ca.* 550 ns, which is very close to the values observed in CsI : Tl.[17] In the case of undoped CLI, a minor fast component of 1.9% (5.5 ns) is present, close to the time resolution of the measurement. The introduction of $In^{+}$ results in the suppression of the fast component in the CLI : In crystal. This effect is even more substantial in the $Tl^{+}$-doped crystal, where the fast component completely vanishes, and the PL decay becomes single exponential. Therefore, we assume the emission centre has a similar origin in all three crystals. $In^{+}$- and $Tl^{+}$-doping enhances carrier trapping, which increases the probability of trapped exciton formation around these impurities.[18] A similar effect was reported for other $Tl^{+}$-doped matrices.[19,20] The fast component is most likely intrinsic and is suppressed *via* the $ns^{2}$ ion doping. A similar fast PL component was observed in undoped CsI.[21]

The DSC heating and cooling curves for the CLI and CLI : In samples are depicted in Fig. 7. Both the heating and cooling DSC curves were repeatedly measured five times for CLI and four times for CLI : In. The onset temperatures and their standard deviations are summarized in Table 5. Both CLI and CLI : In heating curves consisted of two endothermic effects, with average rounded onset temperatures of $T_{onset,1}$ = 190 °C and $T_{onset,2}$ = 213 °C for CLI, and $T_{onset,1}$ = 188 °C and $T_{onset,2}$ = 220 °C for CLI : In. These results indicate congruent melting behaviour, with a eutectic melting point of around 188–190 °C, as can be seen in Fig. 7. The second endothermic effect corresponds to the liquidus temperature.

Using XRPD data (Table 2), we calculated an average pure $Cs_2Li_3I_5$ phase content of roughly 79 wt% within the CLI crystal samples analysed by DSC. In contrast, the XRPD analysis of the $In^{+}$-doped CLI revealed that it was composed entirely of the $Cs_2Li_3I_5$ phase, with only trace amounts of unknown impurities present, as observed by DSC. This is due to the higher sensitivity of the DSC compared to XRPD. As previously discussed in the text, XRPD has relatively low sensitivity (a detection limit of *ca.* 1 wt%). By fitting the two measured values – 213 °C for CLI (assuming 79 wt% purity) and 220 °C for CLI : In (assuming 99 wt% purity) – and presuming linear dependence, we calculated a melting point of approximately 220.4 °C for CLI.

**Table 4** PL decays of CLI, CLI : Tl and CLI : In crystals with the percentage representation of the fast and the slow decay component, and with the $\chi^{2}$ value for each fit

| Crystal | Excitation (nm) | Emission (nm) | Decay time (ns) | | $\chi^{2}$ (−) |
| | | | Fast component | Slow component | |
|---|---|---|---|---|---|
| CLI | 295 | 514 | 5.53 | 523 | 0.9987 |
| | | | 1.9% | 98.1% | |
| CLI : Tl | 295 | 535 | 0 | 557 | 1.1515 |
| | | | 0% | 100.0% | |
| CLI : In | 295 | 528 | 7.59 | 554 | 1.0829 |
| | | | 0.4% | 99.6% | |

**Table 5** Average onset temperatures for CLI and CLI : In for heating and cooling cycles with standard deviation after multiple cycle measurements

| Process | $T_{onset}$ | CLI | CLI : In |
|---|---|---|---|
| Heating (°C) | $T_{onset,1}$ | 189.5 ± 0.1 | 187.7 ± 0.1 |
| | $T_{onset,2}$ | 213.4 ± 0.4 | 220.1 ± 0.3 |
| Cooling (°C) | $T_{onset,1}$ | 168.8 ± 3.7 | 174.6 ± 2.6 |
| | $T_{onset,2}$ | 160.5 ± 1.2 | 170.6 ± 1.3 |

Therefore, the actual melting point of CLI is likely to be close to 220 °C.

The cooling DSC curves exhibited two exothermic effects, with average onset temperatures of $T_{onset,1}$ = 169 °C and $T_{onset,2}$ = 161 °C for CLI, and $T_{onset,1}$ = 174 °C and $T_{onset,2}$ = 171 °C for CLI : In. The first exothermic effect is associated with the crystallisation of the melt, indicating that CLI undergoes approximately 50 °C of undercooling before the crystallisation occurs. The second exothermic effect is attributed to the solidification of the eutectics. Repetitive measurements of CLI and CLI : In demonstrated the reversibility and stability of the observed effects. The high undercooling and low melting point of $Cs_2Li_3I_5$ pose challenges for its crystal growth. Tubular furnaces designed for halide scintillator growth are typically optimised for materials with melting points within the following temperature range: 400–1000 °C. Therefore, there is significant potential to improve crystal quality by optimising the growth conditions. Although the low melting point may seem disadvantageous during the R&D phase, the reduced energy consumption resulting from it could be a significant advantage for commercialisation.

## Conclusions

Ternary $Cs_2Li_3I_5$ is a promising alternative material for combined neutron/gamma sensors due to its high lithium content (30.0 at%), density of 3.93 g cm$^{-3}$, and high effective atomic number ($Z_{eff}$ = 53, calculated according to Murty, R.[22]), offering high detection efficiency for both neutron and gamma radiation, as well as the possibility of producing $^6$Li/$^7$Li-enriched crystals.

In this study, three bulk crystals of CLI, CLI : In, and CLI : Tl were grown using the miniaturised vertical Bridgman method. X-ray powder diffraction analysis confirmed the presence of the $Cs_2Li_3I_5$ ternary phase in all three crystals. The CLI : In crystal was identified as a single phase throughout the ingot, with traces of unknown impurities, as confirmed by the differential scanning calorimetry. Therefore, it is necessary to purify LiI for further experiments and consider a different preparation technique (Synthesis II should be used regardless of the crystal composition). The radioluminescence spectra exhibited broad emission peaks in all crystals, with a redshift observed for CLI : Tl and CLI : In. Specifically, CLI : Tl exhibited maximum emission at around 534 nm, comparable to CsI : Tl, while CLI : In peaked at around 522 nm, comparable to CsI : In. Interestingly, the RL integral intensity of CLI : Tl was ca. 9 times higher than that of the BGO reference sample. The grown crystals exhibited behaviour similar to the CsI matrix; however, photoluminescence measurements indicated that the luminescent centres originate from the CLI matrix, with possible exciton stabilisation around the doped ions. Additional local maxima were identified at approximately 313 and 508 nm for CLI, 380 nm for CLI : Tl, and 320 nm for CLI : In. The primary PL decay time of ca. 550 ns was observed for all three crystals, which is similar to the decay time of CsI : Tl and further supports the impurity-trapped exciton origin. The DSC measurements revealed two endothermic peaks in the heating curves of CLI and CLI : In. The first occurred at ca. 188–190 °C and was attributed to the eutectic point. The second endothermic effect, which occurred at around 220 °C, corresponded to the liquidus temperature. Then, the actual melting point of CLI was estimated to be close to 220 °C. The presence of only two endothermic peaks in the DSC curves implies that CLI melts congruently. Based on the DSC cooling curves, CLI exhibits high undercooling of ca. 50 °C, which is unfavourable for crystal growth.

## Author contributions

Kateřina Křehlíková – conceptualization, data curation, formal analysis, investigation, validation, visualization, writing – original draft, and writing – review & editing. Vojtěch Vaněček – conceptualization, data curation, formal analysis, investigation, validation, visualization, and writing – review & editing. Robert Král – conceptualization, funding acquisition, project administration, investigation, validation, supervision, and writing – review & editing. Romana Kučerková – data curation, formal analysis, investigation, and visualization. Petra Zemenová – data curation, formal analysis, and investigation. Jan Rohlíček – data curation, formal analysis, and investigation. Petr Průša – data curation, formal analysis, investigation, and writing – review & editing. Kateřina Rubešová – conceptualization, data curation, formal analysis, investigation, supervision, and writing – review & editing.

## Conflicts of interest

There are no conflicts to declare.

## Data availability

Data for this article are available at "ASEP – Repository of the Czech Academy of Sciences" at **https://doi.org/10.57680/asep.0636349**.

## Acknowledgements

The work is supported by Operational Programme Johannes Amos Comenius financed by European Structural and Investment Funds and the Czech Ministry of Education, Youth and Sports (Project LASCIMAT – CZ.02.01.01/00/23_020/0008525).